\documentclass[showpacs,amsmath,amssymb,aps,showkeys,floatfix,prd,a4paper]{revtex4}
\usepackage{graphicx} 
\usepackage{dcolumn}
\usepackage{bm}
\usepackage{epsfig}
\usepackage{amsfonts}
\usepackage{amssymb,amscd}
\usepackage{subfigure}
\usepackage{xcolor}
\def\lsim{\raise0.3ex\hbox{$<$\kern-0.75em\raise-1.1ex\hbox{$\sim$}}}

\def\gsim{\raise0.3ex\hbox{$>$\kern-0.75em\raise-1.1ex\hbox{$\sim$}}}

\newcommand{\be}{\begin{equation}}

\newcommand{\ee}{\end{equation}}

\def\beq{\begin{equation}}

\def\eeq{\end{equation}}

\def\beqa{\begin{eqnarray}}

\def\eeqa{\end{eqnarray}}

\newcommand{\ba}{\begin{eqnarray}}

\newcommand{\lb}{\Lambda }

\newcommand{\rr}{\mbox{\boldmath $r$}}

\newcommand{\rb}{\mbox{\boldmath $b$}}

\def\gappeq{\mathrel{\rlap {\raise.5ex\hbox{$>$}}

{\lower.5ex\hbox{$\sim$}}}}

\def\lappeq{\mathrel{\rlap{\raise.5ex\hbox{$<$}}

{\lower.5ex\hbox{$\sim$}}}}

\def\Toprel#1\over#2{\mathrel{\mathop{#2}\limits^{#1}}}

\begin{document}

\begin{flushright}
MS-TP-23-21
\end{flushright}

%%%%%%%%%%%%%%%%%%%%%%%%%%%%%%%%%%

\title{Leading $\Lambda$ production 
in future electron - proton colliders}

%%%%%%%%%%%%%%%%%%%%%%%%%%%%%%%%%%

\author{F. Carvalho}
\affiliation{Departamento de Ci\^encias Exatas e
  da Terra, Universidade Federal de S\~ao Paulo,\\  
  Campus Diadema, Rua Prof. Artur Riedel, 275, Jd. Eldorado,
  09972-270, Diadema, SP, Brazil.}

\author{V.P. Gon\c{c}alves}
\affiliation{Institut f\"ur Theoretische Physik, Westf\"alische Wilhelms-Universit\"at M\"unster,
Wilhelm-Klemm-Stra\ss e 9, D-48149 M\"unster, Germany}
%\affiliation{Institute of Modern Physics, Chinese Academy of Sciences,
%  Lanzhou 730000, China}
\affiliation{Physics and Mathematics Institute, Federal University of Pelotas, \\
  Postal Code 354,  96010-900, Pelotas, RS, Brazil}

\author{K. P. Khemchandani}
\affiliation{Departamento de Ci\^encias Exatas e
  da Terra, Universidade Federal de S\~ao Paulo,\\  
  Campus Diadema, Rua Prof. Artur Riedel, 275, Jd. Eldorado,
  09972-270, Diadema, SP, Brazil.}
  %Universidade Federal de S\~ao Paulo, C.P. 01302-907, S\~ao Paulo, Brazil.

\author{F.S. Navarra}
\affiliation{Instituto de F\'{\i}sica, Universidade de S\~{a}o Paulo,
C.P. 66318,  05315-970 S\~{a}o Paulo, SP, Brazil.}

\author{D. S. Spiering}
\affiliation{Instituto de F\'{\i}sica, Universidade de S\~{a}o Paulo,
C.P. 66318,  05315-970 S\~{a}o Paulo, SP, Brazil.}

\author{A. Mart\'{\i}nez Torres}
\affiliation{Instituto de F\'{\i}sica, Universidade de S\~{a}o Paulo,
C.P. 66318,  05315-970 S\~{a}o Paulo, SP, Brazil.}

%%%%%%%%%%%%%%%%%%%%%%%%%%%%%%%%%%

\begin{abstract}
  Leading $\Lambda$ (LL) production in $ep$ collisions at high energies is      
  investigated using the color dipole formalism and taking into account   
  the nonlinear QCD effects. In particular, the  impact of the absorptive 
  effects on the LL spectra are estimated considering the    kinematical 
  range that will be probed by  the Electron Ion   
  Collider (EIC) and by the Large Hadron electron Collider (LHeC). Our
  results indicate that the LL spectrum is strongly      
  suppressed at small photon virtualities. These results suggest that  
  absorptive  effects are not negligible and should be taken into account in order to extract the kaon structure function from data on  leading
  $\Lambda$ production.
\end{abstract}

%%%%%%%%%%%%%%%%%%%%%%%%%%%%%%%%%%

\pacs{12.38.-t, 24.85.+p, 25.30.-c}
\keywords{Quantum Chromodynamics, Leading Particle Production, Nonlinear QCD 
effects.}

%%%%%%%%%%%%%%%%%%%%%%%%%%%%%%%%%%%

\maketitle

\vspace{0.5cm}

%%%%%%%%%%%%%%%%%%%%%%%%%%%%%%%%%%%

\section{Introduction}

The experimental data from HERA, RHIC and LHC  have  largely improved our    
understanding of the proton structure  during the last decades (For a review 
see, e.g. Ref. \cite{rmp}). In particular, the data from the HERA $ep$   
collider have demonstrated that for high energies and small values of the    
Bjorken - $x$ variable, the proton is composed by a large number of gluons,  
which motivated a large number of theoretical studies about the QCD dynamics 
at high partonic densities \cite{hdqcd}. The observation of nonlinear effects  
in the QCD dynamics is one of the main  motivations for the construction of the  
future electron - hadron colliders at  BNL (EIC) \cite{eic} and at CERN (LHeC) 
\cite{lhec}. In addition, these colliders are expected to probe the 3D          
structure of the proton encoded in the quantum phase space Wigner distributions, 
which include information on both generalized parton distributions (GPDs) and  
transverse momentum parton distributions (TMDs). As a consequence, a deeper 
understanding of the proton structure is expected to be reached in the 
forthcoming years.

A natural question is if a similar improvement of our understanding of the  
pion and kaon structure can also be reached in the near future (See e.g.  
Ref. \cite{pionkaon}). Such a challenge has motivated a large number of     
studies based on the meson cloud model of the proton \cite{Sullivan:1971kd}, 
which have proposed different ways of constraining the meson structure in   
$ep$ and $pp$ colliders  \cite{Holtmann,Kopeliovich:1996iw,Przybycien:1996z,
Nikolaev:1997cn,holt,Kopeliovich:2012fd,McKenney:2015xis,nos1,nos2,
Kumar:2022nqu,Kumar:2022kww,kchina22}. 
The basic idea is that the proton may be decomposed in a series of Fock states, 
containing states such as $|\pi^+n\rangle$ and $|K^+ \Lambda \rangle$, and that  
these states can be probed in the interaction with a given projectile.       
Therefore, the virtual photon present in $ep$ collisions can be used to probe  
the meson structure, with the process being characterized by a leading baryon 
that acts as a spectator carrying a large fraction of the incoming proton 
momentum and having a very large rapidity. In the last decades, this formalism  
has been applied to leading neutron production at HERA 
\cite{Holtmann,Kopeliovich:1996iw,Przybycien:1996z,Nikolaev:1997cn,holt,     
Kopeliovich:2012fd,McKenney:2015xis,nos1,nos2,Kumar:2022nqu,Kumar:2022kww,
kchina22}, which have improved our understanding of the pion structure.      
In particular, in Refs. \cite{nos1,nos2}, the color dipole (CD) formalism, 
which successfully describes the $ep$ HERA data and allows us to take into    
account  the nonlinear QCD dynamical effects, has been extended to 
leading baryon production in inclusive and exclusive processes. As  
demonstrated in Refs. \cite{nos1,nos2}, the CD formalism is able to describe 
the current leading neutron data and important improvements are expected in 
the future $ep$ colliders. One of the goals of this paper is to extend the   
CD formalism to leading $\Lambda$ production and present predictions for the 
EIC and LHeC. Our study is motivated by the analysis performed in Ref. 
\cite{kchina22}, which has demonstrated that a future measurement of this 
process is feasible. 

Another goal is to estimate, for the first time, the impact of the absorptive 
effects on leading $\Lambda$ production in $ep$ collisions. The absorptive     
effects (denoted $S^2_{eik}$ hereafter) are associated with soft rescatterings 
between the produced and spectator particles, which can break the validity
of the factorization hypothesis. This hypothesis allows us to factorize the 
cross section in terms of the $p \rightarrow MB$ splitting  and photon - meson 
interaction. As a consequence, a precise determination of $S^2_{eik}$ is 
fundamental in order to access the meson structure. 
The studies performed in Refs. \cite{pirner,kop,Kho17,Kai06} indicated that 
these effects strongly affect leading neutron production in $pp$ collisions. 
In contrast, the absorptive corrections are predicted to be smaller in $ep$ 
collisions and their effects become weaker at larger photon virtualities    
\cite{pirner,Kho06,levin,nos21}.  In particular, in Ref. \cite{nos21}, one 
has applied the CD formalism to estimate $S^2_{eik}$ in the leading neutron  
production at $ep$ colliders. In this paper, we will extend the analysis 
performed in Ref. \cite{nos21} to the leading $\Lambda$ production and will  
estimate its impact for different center - of - mass energies and different 
photon virtualities. Moreover, a comparison with the results derived for 
leading neutron production will also be presented. The results presented in 
this paper are a necessary step to allow us, in the future, to obtain a 
realistic description of the kaon structure.

This paper is organized as follows. In the next Section we will address  
leading $\Lambda$ production in $ep$ collisions and present its 
description in the color dipole formalism. The main ingredients needed   
for the calculation of the spectrum will be briefly reviewed. Moreover,  
we will discuss the absorptive effects and present our assumptions for the 
dipole - kaon cross section. Our results will be presented in Section     
\ref{res} considering different values of the center - of - mass energy      
and several photon virtualities. A comparison with the results derived for 
leading neutron  production will  be presented. Finally, in Section \ref{sum} 
we will summarize our main results and conclusions.

\begin{figure}[t]
%\begin{tabular}{ccc}
\includegraphics[width=.55\linewidth]{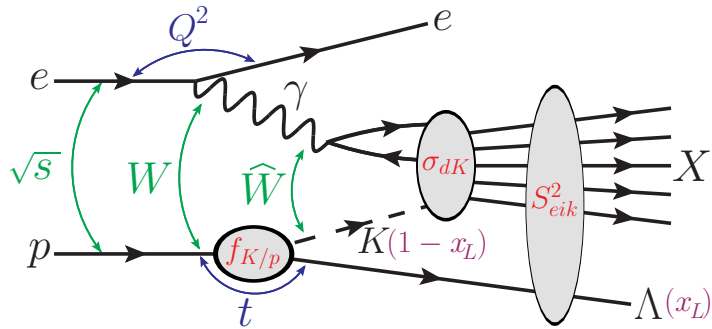}
\caption{Leading  $\Lambda$ production in  
$e p \rightarrow e \Lambda X $  interactions at high energies in the color
dipole model.}
\label{Fig:diagram}
\end{figure}

\section{Leading $\Lambda$ spectrum in the Color Dipole Formalism } 
\label{leading}

The kaon structure can be  
probed in electron - proton collisions through the Sullivan process   
\cite{Sullivan:1971kd}, where the electron scatters off the meson cloud of  
the proton target. The associated processes  can be separated by tagging a 
forward $\Lambda$ in the final state, which carries a large fraction of the   
proton energy. Theoretically, this leading $\Lambda$ production is usually  
described assuming that the splitting  $p \rightarrow \Lambda  K$  and  the 
photon -- kaon interaction  can be factorized, as represented in 
Fig.~\ref{Fig:diagram}, where $f_{K/p}$ represents the kaon flux. Assuming  
the validity of the factorization hypothesis and the universality of the  
fragmentation process, which allows us to constrain $f_{K/p}$ using the 
data of leading $\Lambda$ production in $pp$ collisions, we can obtain    
$\sigma_{\gamma^* K}$ and, consequently,  determine the $x$ and $Q^2$       
dependence of the kaon structure function. Following Refs. \cite{nos1,nos2} 
we propose  to treat  leading $\Lambda$   
production in  $ep$ processes with the color dipole  formalism. 
In this model, the virtual photon - kaon cross section can be factorized in 
terms of the photon wave function (which describes the photon splitting in    
a $q\bar{q}$ pair) and the dipole - kaon cross section $\sigma_{d K}$ 
(See Fig. \ref{Fig:diagram}).  As shown in Refs. \cite{nos1,nos2},   
the HERA data on leading neutron production are quite well described by this     
approach assuming that absorptive corrections can be factorized  and represented 
by a multiplicative constant factor, denoted by $K_{abs}$ in Ref. \cite{nos1}. 

Let us first recall the expressions proposed in Ref. \cite{nos1} to treat    
leading neutron production in $ep$ collisions, adapting them to leading 
$\Lambda$ production. Disregarding initially the absorptive effects 
($S^2_{eik}$ = 1), this process  can be seen as a set of three factorizable 
subprocesses:  i) the photon emitted by the electron  
fluctuates into a quark-antiquark pair (the color dipole), ii) the color dipole 
interacts with the kaon  and iii) the leading $\Lambda$ is formed. In the color 
dipole formalism, the differential cross section reads: 
\begin{eqnarray}
  \frac{d^2 \sigma(W,Q^2,x_L)}{d x_L} & = & \int dt \, f_{K/p} (x_L,t) \, \times
  \sigma_{\gamma^* K}(\hat{W}^2,Q^2) \,\,, \\
& = & \int dt \, f_{K/p} (x_L,t) \times \int _0 ^1 dz \int d^2     
\rr \sum_{L,T} \left|\Psi_{T,L} (z, \rr, Q^2)\right|^2  
\sigma_{dK}({x}_{K}, \rr)
\label{crossgen}
\end{eqnarray}
where $Q^2$ is the virtuality of the exchanged photon, $x_L$ is the proton  
momentum fraction carried by the $\Lambda$ and $t$ is the square of the 
four-momentum of the exchanged kaon. Moreover,  $\hat{W}$ is the center-of-mass 
energy of the virtual photon-kaon system, which can be written as 
$\hat{W}^2 = (1-x_L) \, W^2$, where $W$ is the center-of-mass energy of the 
virtual photon-proton system.  In terms of the measured quantities $x_L$ and 
transverse momentum $p_T$ of the $\Lambda$, the kaon virtuality is:
\beq
t \simeq-\frac{p_T^2}{x_L}-\frac{(1-x_L)(m_\lb^2-m_p^2 x_L)}{x_L} \,\,.
\label{virtuality}
\eeq
In Eq. (\ref{crossgen}), the virtual photon - kaon cross section was expressed 
in terms  of the transverse and longitudinal photon wave functions $\Psi_i$,  
which describe the photon splitting into  a $q\bar{q}$ pair of size           
$r \equiv |\rr|$, and the dipole-kaon cross section  $\sigma_{d K}$, which is  
determined by the QCD dynamics at high energies \cite{hdqcd}.  The variable $z$ 
represents  the longitudinal photon momentum fraction carried by the quark, the 
variable $\rr$  defines the relative transverse separation of the pair (dipole) 
and the scaling variable $x_{K}$ is defined by $x_{K} = x / (1-x_L)$, where $x$ 
is the Bjorken variable. As in Ref. \cite{nos1}, we will assume that this 
quantity can be related to the dipole - proton cross section using the additive 
quark model. Moreover, $\sigma_{dp}$ will be  described by the Color Glass     
Condensate (CGC) formalism, as given in the phenomenological model proposed in 
Ref. \cite{iim}. As a consequence, we will have that:
\begin{eqnarray}
\sigma_{d K} (x,\rr) =  \frac{2}{3} \cdot \sigma_{dp} ({x}, \rr) =
\frac{2}{3} \cdot  2 \pi R_p^2  \times  \left\{ \begin{array}{ll}
{\mathcal N}_0\, \left(\frac{r\, Q_s}{2}\right)^{2\left(\gamma_s + 
\frac{\ln (2/r\, Q_s)}{K \,\lambda \, Y}\right)}\,, & 
\mbox{for $rQ_s({x}) \le 2$}\,,\\
1-\text{e}^{-a\,\ln^2\,(b\,r\, Q_s)}\,,  & \mbox{for $r Q_s({x})  > 2$}\,,
\end{array} \right.
\label{Eq:sigdp}
\end{eqnarray}
where  $a$ and $b$ are determined by continuity conditions at $r Q_s({x})=2$.
The parameters $\gamma_s= 0.7376$, $\kappa= 9.9$, ${\mathcal N}_0=0.7$ and
$R_p = 3.344$ GeV$^{-1}$ have been adjusted using the HERA data in
Ref. \cite{soyez}, with the saturation scale $Q_s$ given by:
\beq
Q^2_s ({x}) = Q^2_0 \left( \frac{x_0}{{x}}\right)^{\lambda}
\label{qsat}
\eeq
with $x_0=1.632\times 10^{-5}$, $\lambda=0.2197$, $Q_0^2 = 1.0$ GeV$^2$.
The first line of Eq. (\ref{Eq:sigdp}) describes the linear regime whereas
the second one includes saturation effects.

The flux factor $f_{K/p}$  gives the probability of the splitting of a proton 
into a $K - \Lambda$ system and can be expressed as follows (See e.g. Ref.   
\cite{pirner}) 
\beq
f_{K/p}(x_L,t) = \frac{1}{2}\pi\sum_{\lambda\lambda'}
        |\phi_{\lb K}^{\lambda\lambda'}(x_L,{\bf p}_T)|^2
\eeq
where $\phi_{\lb K}^{\lambda\lambda'}(x_L,{\bf p}_T)$ is the probability
amplitude to find, inside a proton with spin up, a $\Lambda$ with longitudinal
momentum fraction $x_L$, transverse momentum ${\bf p}_T$ and helicity 
$\lambda$ and a kaon, with longitudinal momentum fraction $1-x_L$,           
transverse momentum $-{\bf p}_T$ and helicity $\lambda'$. In the light-cone  
approach, the amplitudes $\phi_{\lb K}$ of a proton with spin $+1/2$, read:
\beqa
\label{phi}
\phi_{\lb K}^{1/2,0}(x_L,{\bf p}_T) & = & 
          \frac{g_0}{4 \pi} \sqrt{\frac{3}{2 \pi}}\frac{1}{\sqrt{x_L^2(1-x_L)}}
          \frac{(x_L m_{N} -m_{\lb})}{M_{\lb K}^2-m_{N}^2}\nonumber\\
\phi_{\lb K}^{-1/2,0}(x_L,{\bf p}_T) & = & 
\frac{g_0}{4 \pi} \sqrt{\frac{3}{2 \pi}}
\frac{1}{\sqrt{x_L^2(1-x_L)}}
          \frac{|{\bf p}_T|e^{-i\varphi}}{M_{\lb K}^2-m_{N}^2}\,,
\eeqa
where $M_{\lb K}^2$ is the invariant mass of the $K - \Lambda$ system,
given by
\[
M_{\lb K}^2 = \frac{m_{\lb}^2+p_T^2}{x_L} + \frac{m_K^2+p_T^2}{1-x_L}\,,
\]
with  $m_\lb$ and $m_K$ being the $\Lambda$ and the kaon masses, $g_0$ is 
the $N K \Lambda$ coupling constant \cite{bnn99} and $\varphi$ is the azimuthal 
angle in the transverse plane. Because of the extended nature of the hadrons 
involved, the interaction amplitudes in the above equations have to be          
modified by including a phenomenological $N K \lb $ form factor, 
$G(x_L,p_T)$. It is important to stress here that while the vertex is
derived from an effective meson-nucleon Lagrangian, the
form factor is introduced ad hoc. In our analysis we will choose the
covariant form factor, corrected by the Regge factor, given by 
\beq
\label{covff}
G(x_L,p_T)  =  {\rm exp}[R_{c}^2(t-m_K^2)] \, (1-x_L)^{- \alpha_0  t}
\eeq
where $\alpha_0 = 1$ GeV$^{-2}$ and  $R_c^2 = 0.3$ GeV$^{-2}$ were 
constrained using the HERA data (For details see Ref. \cite{nos1}). 
The amplitude 
$\phi_{\lb K}^{\lambda\lambda'}(x_L,{\bf p}_T)$  changes to
$\phi_{\lb K}^{\lambda\lambda'}(x_L,{\bf p}_T) \, G(x_L,p_T)$ and then the
kaon flux becomes: 
\beq
\label{flux}
f_{K/p}(x_L,t) = \frac{1}{2}\pi\sum_{\lambda\lambda'}
        |\phi_{\lb K}^{\lambda\lambda'}(x_L,{\bf p}_T)|^2|G(x_L,p_T)|^2\,,
\eeq
where $1/2$ is the isospin factor and the azimuthal angle in the transverse 
plane has been integrated out.

In order to derive more realistic predictions for the leading $\Lambda$    
spectrum it is crucial to improve the description of $S^2_{eik}$. This was 
done in Ref.~\cite{nos21} for leading neutron production, where we revisited  
and updated the approach proposed in Ref. \cite{pirner} to study absorptive 
effects. In order to include the absorptive effects  in our predictions for  
the leading $\Lambda$  spectrum $d\sigma/dx_L$, we will follow the approach  
proposed in Ref. \cite{pirner}, where these effects were estimated using the 
high - energy Glauber approximation \cite{glauber} to treat the multiple   
scatterings between the dipole and the $K-\Lambda$ system. As demonstrated   
in Ref. \cite{pirner}, such approach can be easily implemented in the impact 
parameter space, implying that the spectrum can be expressed as follows
\beqa
\label{desdzgamma}
\frac{d\sigma(W,Q^2,x_L)}{dx_L} & = & 
        \int\! d^2{\rb}_{rel} \, \rho_{\lb K}(x_L,{\rb}_{rel})\,
        \int\! dz \, d^2{\rr} \,\sum_{L,T} \left|\Psi_{T,L} 
(z, \rr, Q^2)\right|^2  \sigma_{d K}({x}_{K}, \rr) \,
        S_{eik}^2(\rr,\rb_{rel}) \,\,\,,
\eeqa
where $ \rho_{\lb K}(x_L,{\rb}_{rel})$ is the probability density of         
finding a $\Lambda$ and a kaon with momenta $x_L$ and $1-x_L$, respectively, 
and with a relative transverse separation $\rb_{rel}$, which is given by  
\beq
\rho_{\lb K}(x_L,{\rb}_{rel}) = \sum_i|\psi^i_{\lb K}(x_L,{\rb}_{rel})|^2\,.
\label{rho}
\eeq
with 
\beq
\psi^i_{\lb K}(x_L,{\rb}_{rel}) = \frac{1}{2\pi}\int\!d^2{\bf p}_T \,  
        e^{i{\rb}_{rel}\cdot{\bf p}_T} \, \phi^i_{\lb K}(x_L,{\bf p}_T) \,,
\eeq
and $\phi^i_{\lb K}$ = $\sqrt{2/3} \, \phi^{\lambda\lambda'}_{\lb K} \,  
G(x_L,p_T)$.  Moreover, the survival factor $S_{eik}^2$ associated to the 
absorptive effects is expressed in terms of the  dipole -- $\Lambda$ 
($\sigma_{d\lb}$) cross section as follows 
\begin{eqnarray}
S_{eik}^2(\rr,\rb_{rel}) = \Big\{1-\Lambda_{\rm eff}^2\frac{\sigma_{d\lb}
(x_\lb,{\rr})}{2\pi}\,
    {\rm exp}\Big[-\frac{\Lambda_{\rm eff}^2{\rb}_{rel}^2}{2}\Big]\Big\}\,,
\label{Eq:esse2}    
\end{eqnarray}
where $x_\lb = x/x_L$ and $\Lambda^2_{\rm eff}$ is an effective parameter   
that was found to be equal to $0.1$ GeV$^2$ in Ref. \cite{pirner}. 

As it can be seen from the above expression, the absorptive corrections depend 
crucially on the dipole - Lambda cross section, $\sigma_{d\lb}$. In our previous 
work \cite{nos21} on the leading neutron spectrum, we assumed that 
$\sigma_{d n}$ was equal to the dipole - proton cross section, $\sigma_{dp}$, 
which is strongly constrained by  HERA data. While this was a good approximation
for $\sigma_{d n}$, this may be not so good for $\sigma_{d\lb}$. What is the 
effect of changing a light quark by a strange quark in the baryon which 
scatters against the dipole ?  We can try to answer this question comparing the
data on $p p$ and $\Sigma^- p$ cross sections measured at $\sqrt{s} \simeq 30$ 
GeV. This energy is relevant for the study of HERA data,  where the 
dipole-baryon collisions happen at $\hat{W} = \sqrt{1-x_L} W$. Considering 
values of $W$ in the range $100 < W < 200$ GeV and assuming that 
$x_L \simeq 0.7$ the dipole-baryon collision energies are typically $50 -100$ 
GeV. At these energies the data and the parametrizations performed in 
Ref. \cite{luna} yield:
$\sigma_{p \Sigma^-} \simeq 0.8 \, \sigma_{pp}$. This suggests that, in a first  
approximation, $\sigma_{d\lb} \simeq 0.8 \, \sigma_{dn}$. Going from HERA to the 
LHeC, $\hat{W}$ would be much higher ($\simeq 500$ GeV) and according to      
\cite{luna} the differences between the cross sections would decrease. On the 
other hand, going from HERA to JLab we have $\hat{W} = 5 - 10$ GeV. In order 
to have an estimate of the dipole - baryon cross section at low energies, in 
the Appendix \ref{MB} we have computed the $\pi - n $ and $\pi - \Lambda$ cross 
sections using a  effective Lagrangian theory \cite{beto1,beto2} and found that it 
predicts that strange baryons have smaller cross section than the light 
baryons.  These findings are in agreement with old data \cite{ldata} and also 
with other calculations \cite{lambes}, indicating that 
$\sigma_{p \Lambda} \simeq 0.8 \, \sigma_{pp}$.                            
Taken together, these results are a strong indication that 
$\sigma_{d\lb}= r_s \sigma_{dp}$, with $r_s \approx cte \le 1$. As we will  
demonstrate in the next Section, when such an assumption is applied to our 
study, it has the interesting consequence that leading $\Lambda$'s have a 
smaller production cross section but are less absorbed than leading neutrons. 
\begin{figure}
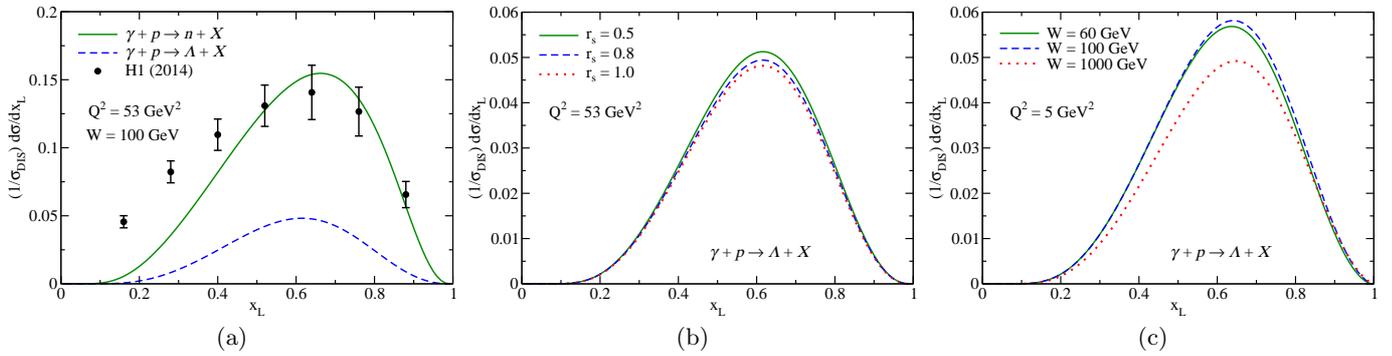

\begin{tabular}{ccc}
 \includegraphics[width=.33\linewidth]{lela-fig2a.eps}&
 \includegraphics[width=.33\linewidth]{lela-fig2b.eps}&
 \includegraphics[width=.33\linewidth]{lela-fig2c.eps} \\
  (a) & (b) & (c)
   \end{tabular}
\caption{(a) Predictions of the color dipole formalism for  leading 
$\Lambda$ production in the HERA kinematical range. For comparison,     
the results for  leading neutron production and corresponding HERA data 
\cite{hera1} are also presented.
(b)  Dependence on $r_s$ of the leading $\Lambda$ spectrum. 
(c) Dependence of the  leading $\Lambda$ spectrum on the photon - proton 
center - of - mass energy $W$ for  $Q^2 = 5$ GeV$^2$.}
\label{Fig:comp}
\end{figure}

\section{Results}
\label{res} 

In this Section we will present the color dipole model predictions for 
leading $\Lambda$ production in $ep$ collisions at the EIC and LHeC. 
Initially, in Fig. \ref{Fig:comp} (a)  we present our prediction for the 
spectrum in the HERA kinematical region,  assuming that $Q^2 = 53$ GeV$^2$, 
$W = 100$ GeV and $r_s = 0.8$. For comparison, we also present the results  
derived in Ref. \cite{nos1} and the associated H1 data \cite{hera1}. It is  
important to emphasize that our predictions are expected to be valid in the 
region $x_L \gtrsim 0.5$, since for smaller values of $x_L$, additional  
contributions are expected to play a significant role  \cite{kop,Kai06}.   
The cross section for leading $\Lambda$ production is smaller than that        
for  leading neutron production, with the peak occuring for $x_L \approx 0.6$. 
In Fig. \ref{Fig:comp} (b) we analyze the dependence of our prediction on the 
value of $r_s$, which is the scale factor that determines the relation between 
the dipole - Lambda and dipole - proton cross sections. As it can be seen, 
larger values of $r_s$ imply a reduction of the spectrum, which is expected    
since the impact of the absorptive effects are larger when $\sigma_{d\Lambda}$ 
increases [See Eq. (\ref{Eq:esse2})].                                      
In Fig. \ref{Fig:comp} (c) we  present our predictions for the  $\Lambda$  
spectra at different center - of - mass energies, $Q^2 = 5$ GeV$^2$ and      
$r_s = 0.8$. We find that the predictions are not strongly dependent on $W$, 
similarly to what was observed in leading neutron production in Ref.          
\cite{nos1}. Such results demonstrate that the color dipole formalism predicts 
that the leading baryon spectrum leads to Feynman scaling, i.e. the energy 
independence of the $x_L$ spectra.
\begin{figure}[t]
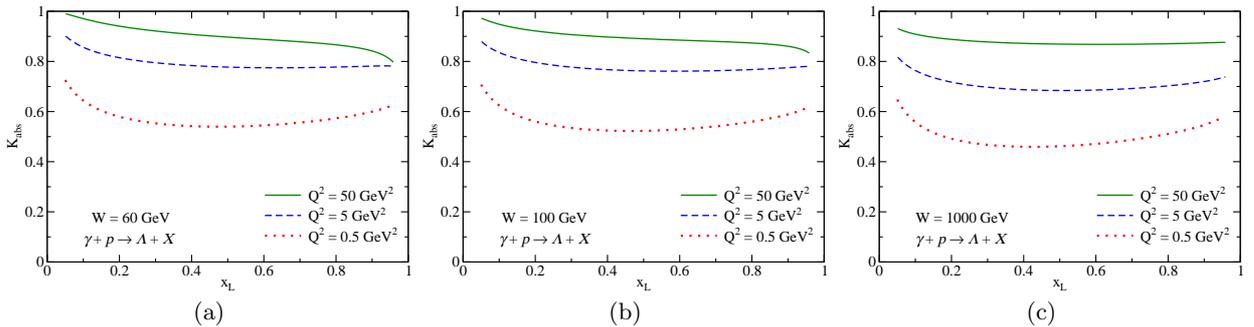

\begin{tabular}{ccc}
  \includegraphics[scale=0.22]{lela-fig3a.eps} &
{\includegraphics[scale=.22]{lela-fig3b.eps}} &
  \includegraphics[scale=0.22]{lela-fig3c.eps} \\
  (a) & (b) & (c)
   \end{tabular}
\caption{Dependence of the absorptive effects on $x_L$ in leading $\Lambda$ 
production in $ep$ collisions for differents values of the photon virtuality
and a) $W = 60$ GeV;  b) $W = 100$ GeV and c) $W = 1000$ GeV.}
\label{Fig:kfac}
\end{figure}

As discussed in the previous Section, in order to measure the $\gamma \pi$ 
and $\gamma K$ cross sections and extract, respectively, the pion and kaon   
structure functions, it is crucial to have control of the absorptive effects. 
In particular, we should know the dependence of these effects on $Q^2$, $W$ 
and $x_L$. We can estimate the impact of the absorptive effects through the     
calculation of the ratio between the cross sections with and without absorption, 
defined by
\begin{eqnarray}
K_{abs}(W,Q^2,x_L) = \frac{\frac{d\sigma}{dx_L}[S^2_{eik}]}
{\frac{d\sigma}{dx_L}[S^2_{eik} = 1]} \,\,.
\end{eqnarray}
In the case of leading $\Lambda$ production, our predictions for this ratio, 
obtained assuming $r_s = 0.8$, are presented in Fig. \ref{Fig:kfac}. Our 
results show that the impact increases for smaller 
values of $Q^2$ and it is almost insensitive to the energy $W$. For  
$Q^2 = 50$ GeV$^2$, we see that  $K_{abs} \approx 0.9$ for $x_L \gtrsim 0.5$, 
with the predictions being similar for the three values of $W$.  
This weak absorption is expected in the color dipole formalism, since at     
large values of $Q^2$ the main contribution to the cross section comes from 
dipoles with a small pair separation. In this regime, known as  color      
transparency, the impact of the rescatterings is small, which implies that  
the absorptive effects become negligible. Another important aspect, is that  
for large photon virtualities, the main effect of absorption is to suppress  
the cross section by an almost constant factor. Similar results were derived 
in Refs. \cite{pirner,nos21}. On the other hand, for small virtualities 
($Q^2 = 0.5$ GeV$^2$), we observe strong absorptive effects, which reduce the 
cross sections by a factor $\approx 0.5$ for $x_L = 0.5$. This result is
also expected, since for small $Q^2$ the cross section is dominated by large
dipoles and, consequently, the contribution of the rescatterings cannot be
disregarded. For larger values of $x_L$, absorptive effects cannot be
modelled by a constant factor. Our conclusions agree with those derived in   
Ref. \cite{Kai06} using Regge theory. Finally, our results     
indicate that the contribution of the absorptive effects is not strongly
energy dependent. This result suggests that the main conclusion of Ref. 
\cite{nos1}, that the spectra will satisfy Feynman scaling, is still valid 
when the absorptive effects are estimated using a more realistic model.

\begin{figure}[t]
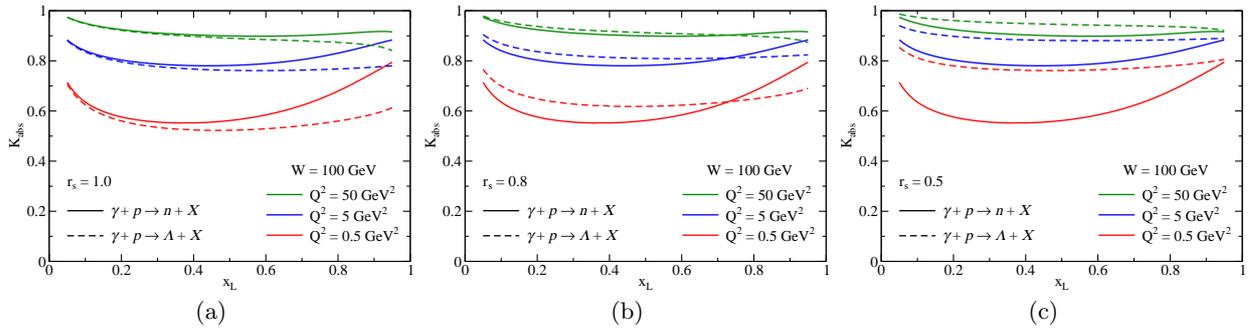

\begin{tabular}{ccc}
  \includegraphics[scale=0.22]{lela-fig4a.eps} &
{\includegraphics[scale=.22]{lela-fig4b.eps}} &
  \includegraphics[scale=0.22]{lela-fig4c.eps} \\
  (a) & (b) & (c)
   \end{tabular}
\caption{Comparison of the  absorptive effects in leading neutron and 
leading $\Lambda$ production in $ep$ collisions at the EIC energy 
($W = 100$ GeV) for different photon virtualities $Q^2$ and different 
values of $r_s$: (a) $r_s = 1.0$  ; (b) $r_s = 0.8$  and (c) $r_s = 0.5$.}
\label{fig-rs}
\end{figure}

Finally, in Fig.~\ref{fig-rs} we present a comparison between the 
absorptive factors estimated for the leading neutron and leading $\Lambda$ productions at the EIC energy ($W = 100$ GeV) considering different values of  $Q^2$. Moreover, the leading $\Lambda$ predictions are presented for three distinct values of $r_s$.  As expected from the results discussed above, these predictions are strongly dependent on $r_s$, with $K_{abs}^{\Lambda} \ge K_{abs}^{n}$ for $r_s \lesssim 0.8$ and intermediated values of $x_L$, indicating that the $\Lambda$'s are less absorbed in comparison to leading neutrons. Moreover, our results indicate that  taking $K_{abs}$ as a constant is, in general, a good approximation for the leading $\Lambda$ production, which is  is useful for practical calculations and for a future determination of the kaon structure.  

\section{Summary}
\label{sum}

In this paper we have investigated  leading $\Lambda$ production in future 
$ep$ colliders using the color dipole formalism and estimated, for the first 
time, the impact of the absorptive corrections in the associated spectrum.  
Our analysis has been motivated by the perspective of using this process to 
measure the kaon structure function and improve our understanding of the 
partonic structure of this meson. We have presented predictions for the 
kinematical ranges that will be probed by the future EIC and LHeC. Our results 
indicate that  leading $\Lambda$  spectra are not strongly energy dependent at 
small photon virtualities. Moreover, we have estimated the impact of the 
absorptive effects, demonstrated that they increase at smaller photon
virtualities and that they depend on the longitudinal momentum $x_L$. 
Our  results show that  modelling  these effects by a constant factor is a
good approximation only for large $Q^2$. Our main conclusion is that a 
realistic measurement of the  $\gamma K$ cross sections in 
future colliders and the extraction of the kaon structure function  must take 
into account the important contribution of the absorptive effects. Future    
experimental data on leading $\Lambda$  production in $ep$ collisions at the 
EIC will be crucial to test the main assumptions of our model, as well as to 
improve our understanding of this important observable.

\begin{acknowledgments}
This work was  partially financed by the Brazilian funding
agencies CNPq, CAPES, FAPESP,  FAPERGS and INCT-FNA (process number 
464898/2014-5). K.P.K and A.M.T gratefully acknowledge the  support from the Funda\c c\~ao de Amparo \`a Pesquisa do Estado de S\~ao Paulo (FAPESP), processos n${}^\circ$  2022/08347-9 and 2023/01182-7.

 \end{acknowledgments}

\appendix
\section{Meson - Baryon cross sections}
\label{MB}

As it was mentioned in the main text, it is very important to have a 
good estimate of the dipole-Lambda cross section. We need to impose 
constraints on this quantity. The first one is that it should be of 
the same order of magnitude as that of the dipole-proton cross section, which is
well known. However this may not be enough, since factors of two or three 
may significantly change the predictions of the absorption factor. Moreover
it is crucial to make predictions for higher energies (for LHeC) and also
for lower energies (JLAB). 
In this Appendix we discuss the meson-baryon cross sections at lower energies. 
In particular we estimate the $\pi$-nucleon and $\pi$-hyperon cross sections. 
The comparison of these cross sections should give (or not) support to the 
choice of the factor $r_s$ used in the main text.

To determine the relation between the $\pi N$ and $\pi\Lambda$ cross sections we follow Refs.~\cite{beto1,beto2}, which consider effective Lagrangians based on the hidden local symmetry approach~\cite{Bando:1985rf} and treats pseudoscalar-baryon an vector-baryons as coupled channels. Using effective Lagrangians for the vector-pseudoscalar-pseudoscalar, vector-vector-vector and vector-baryon-baryon vertices, $t-$, $u-$, $s-$channels contributions as well as contact terms were considered in Refs.~\cite{beto1,beto2} to determine the amplitudes $V_{ij}$ describing the transitions $PB\to PB$, $VB\to VB$ and $PB\to VB$. These amplitudes are projected on the $s$-wave, isospin and spins $1/2$ and $3/2$. Using the latter amplitudes, the Bethe-Salpeter equation,
\begin{align}
T=V+VGT,\label{BS}
\end{align}
is then solved in a coupled-channel approach, with $V$ and $G$ being matrices in the coupled channel space. The elements of $V$ are the amplitudes $V_{ij}$ describing the transition from a channel $i$ to a channel $j$ formed by pseudoscalar-baryon (PB) or vector-baryon particles (VB). The $G$ in Eq.~(\ref{BS}) has, as elements, the corresponding $PB$ and $VB$ loop functions, which are regularized with either a cut-off of $\sim 600$ MeV or dimensional regularization. In particular,
\begin{align}
G_k=i 2M_k\int\frac{d^4q}{(2\pi)^4}\frac{1}{(p-q)^2-M^2_k+i\epsilon}\frac{1}{q^2-m^2_k+i\epsilon},
\end{align}
where $p$ is the total four-momenta, $M_k$ ($m_k$) is the baryon (meson) mass of the channel $k$. We refer the reader to Refs.~\cite{beto1,beto2} for the precise expressions of the elements of $V$. 

By solving Eq.~(\ref{BS}), the two-body $T-$matrix for the system is determined and poles in the complex energy plane which can be associated with known states (like $N^*(1535)$, $N^*(1650)$, $\Lambda(1405)$) are found. The parameters of the model are constrained to reproduce the data on cross sections like $K^-p\to K^-p$, $\bar K^0n$, $\eta\Lambda$, $\pi^0\Lambda$, $\pi^0\Sigma^0$, $\pi^\pm\Sigma^\mp$. The CLAS data for $\gamma p\to K^+\Lambda(1405)$ are also well reproduced~\cite{Kim:2021wov}. For strangeness $-1$, channels like $\bar K N$, $\pi\Sigma$, $\pi\Lambda$, $\eta\Sigma$, $K\Xi$, $\bar K^* N$, $\rho\Lambda$, $\rho\Sigma$, $\omega\Sigma$, $K^*\Xi$, $\phi\Sigma$ were considered as coupled in Ref.~\cite{beto2}. In this way, a transition like $\pi\Lambda\to\pi\Lambda$ can have contributions from, for example, a virtual $\pi\Sigma$ channel. For strangeness 0, channels like $\pi N$, $\eta N$, $K\Lambda$, $K\Sigma$, $\rho N$, $\omega N$, $\phi N$, $K^*\Lambda$, $K^*\Sigma$ are treated as coupled channels when solving Eq.~(\ref{BS})~\cite{beto1}. 

\begin{figure}[h!]
\begin{tabular}{ccc}
 \includegraphics[width=.38\linewidth]{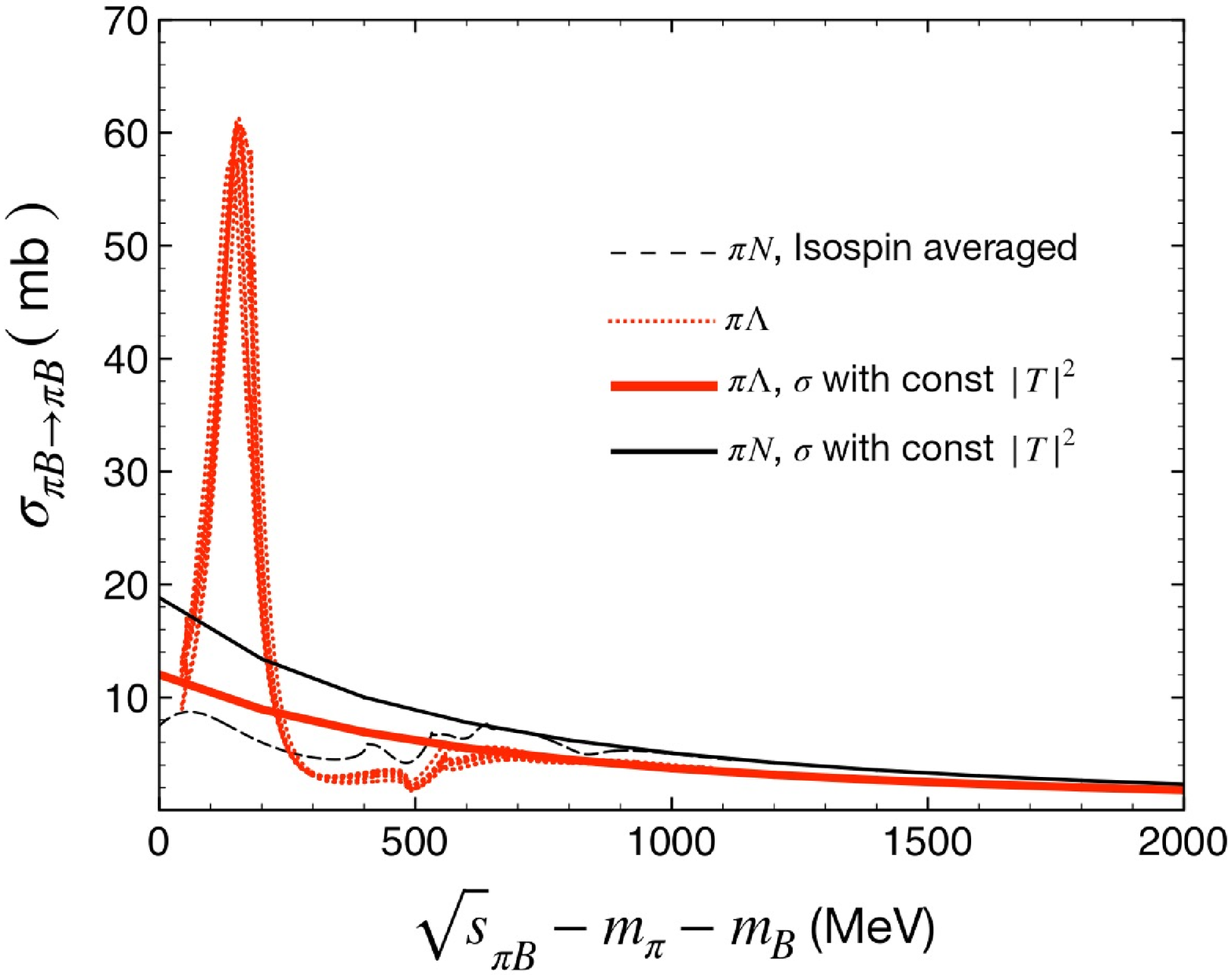} & \,\,\, &
\includegraphics[width=.38\linewidth]{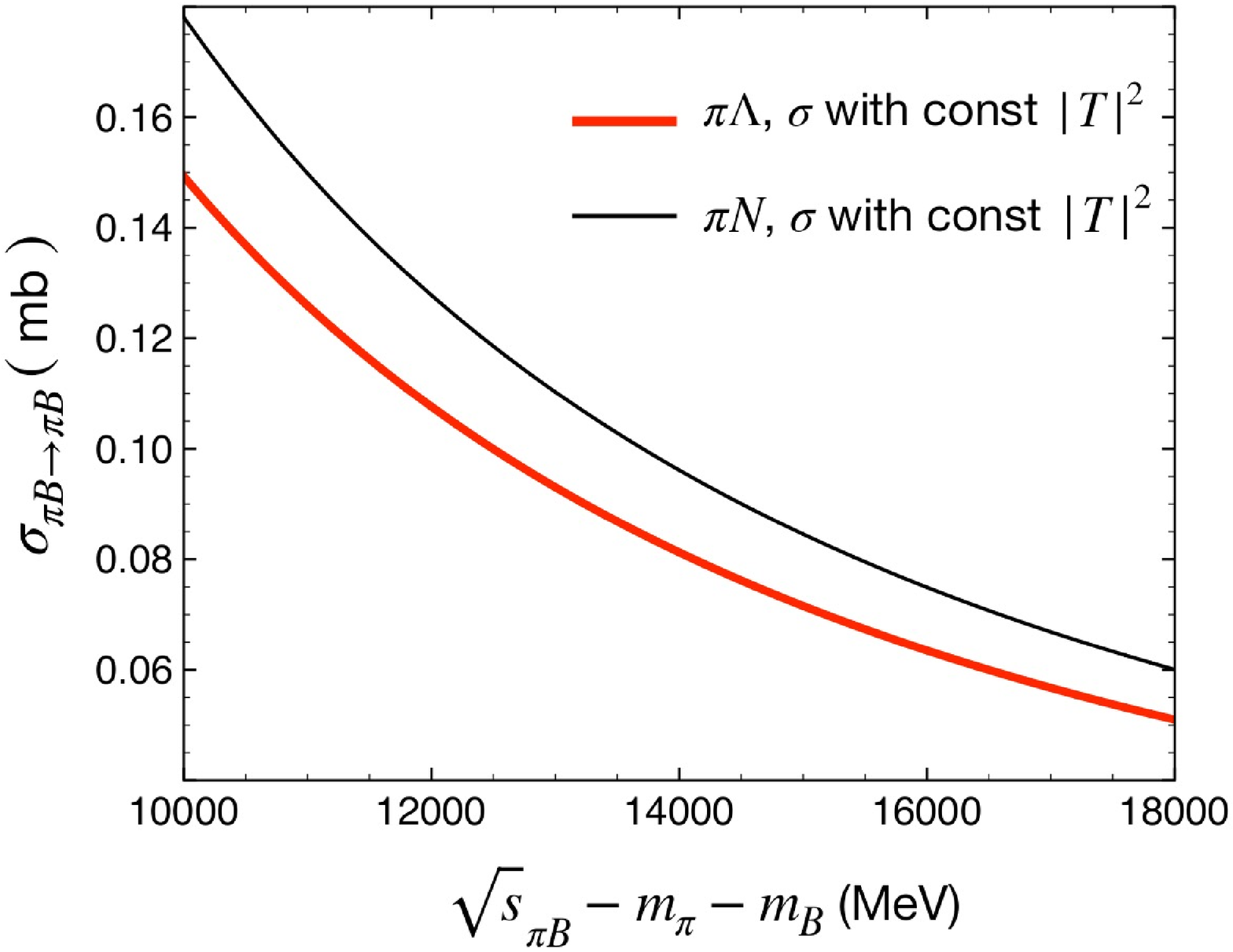} \\
  (a) & \,\,\, & (b)
   \end{tabular}
\caption{$\pi - \text{baryon}$ cross sections. a) Lower energies 
b) Extrapolation to higher energies. }
\label{sigs}
\end{figure}

In Fig.~\ref{sigs} we show the results for the $\pi N\to \pi N$ and $\pi\Lambda\to \pi\Lambda$ cross sections. The results shown as dashed and dotted lines, in the  Fig.~\ref{sigs}a, correspond to the $\pi N$ and $\pi \Lambda$ cross sections, respectively, obtained by using the amplitudes of Refs.~\cite{beto1,beto2}.
The total  spin is $1/2$ in both cases, while the  cross sections shown for the $\pi N\to \pi N$ are isospin averaged. It is important to mention here that the $\pi \Lambda$ cross sections   correspond to a set of  curves since all of them lead to similar quality of fits in Ref.~\cite{beto2}. Though some structures are seen at energies near the threshold, which correspond to the appearance of resonances, the cross sections for excitation energies ($\sqrt{s}_{\pi B}-m_\pi-m_B$) above 1000 MeV attain a smooth  dependence in both $\pi N$ and $\pi \Lambda$ cases. It seems then reasonable, for the purpose of an estimation of the cross section ratios at higher energies, to extrapolate the results to such energies by considering the corresponding $T$-matrices to remain constant beyond some point.

 We show, in Fig.~\ref{sigs}, the cross sections obtained by considering a constant amplitude multiplied to the phase space as solid lines.   The former is the value of the amplitude determined at the  highest energy in Refs.~\cite{beto1, beto2}. As can be seen from the Fig.~\ref{sigs}a, such cross sections (obtained with a constant amplitude) do not describe the structures at low energies (shown by dashed lines) but do completely agree with the results of Refs.~\cite{beto1,beto2} beyond the excitation energies of 1000 MeV. 

Thus, basically the energy dependence of the $\pi N$ and $\pi \Lambda$ cross sections follows the corresponding phase space. Such behavior, if extrapolated to higher energy regions (see Fig.~\ref{sigs}b), can provide an estimation of the ratio between the cross sections of $\pi N$ and $\pi \Lambda$, yielding 
$\sigma_{\pi\Lambda} \simeq  0.83 \, \sigma_{\pi N}$ 
for $\sqrt{s}\simeq 11-19$ GeV.

\end{document}